# Operationalising AI governance through ethics-based auditing: An industry case study

Jakob Mökander[1] and Luciano Floridi[1,2]

[1] Oxford Internet Institute, University of Oxford, 1 St Giles', Oxford, OX1 3JS, UK

[2] Department of Legal Studies, University of Bologna, Via Zamboni 33, Bologna, 40126, Italy

Correspondence author: Jakob Mokander <jakob.mokander@oii.ox.ac.uk>

**Abstract**

Ethics-based auditing (EBA) is a structured process whereby an entity's past or present behaviour is assessed for consistency with moral principles or norms. Recently, EBA has attracted much attention as a governance mechanism that may bridge the gap between principles and practice in AI ethics. However, important aspects of EBA – such as the feasibility and effectiveness of different auditing procedures – have yet to be substantiated by empirical research. In this article, we address this knowledge gap by providing insights from a longitudinal industry case study. Over 12 months, we observed and analysed the internal activities of AstraZeneca, a biopharmaceutical company, as it prepared for and underwent an ethics-based AI audit. While previous literature concerning EBA has focused on proposing evaluation metrics or visualisation techniques, our findings suggest that the main difficulties large multinational organisations face when conducting EBA mirror classical governance challenges. These include ensuring harmonised standards across decentralised organisations, demarcating the scope of the audit, driving internal communication and change management, and measuring actual outcomes. The case study presented in this article contributes to the existing literature by providing a detailed description of the organisational context in which EBA procedures must be integrated to be feasible and effective.

**Keywords**

artificial intelligence; auditing; case study; ethics; governance; industry; practice

## 1. Introduction

Recent publications have identified ethics-based auditing (EBA) as a governance mechanism with the potential to help bridge the gap between principles and practice in 'AI ethics' (see e.g., Brown et al., 2021a; Brundage et al., 2020a; Koshiyama et al., 2021; Mökander & Floridi 2021; Raji et al., 2019). Simplified, EBA is a structured process whereby an entity's present or past behaviour is assessed for consistency with relevant principles or norms.[1] The promise of EBA is underpinned by two ideas. The first is that procedural regularity and transparency contribute to good governance (Floridi, 2017; Larsson, 2020). The second is that proactivity in the design of AI systems help identify risks and prevent harm before it occurs (Kazim & Koshiyama, 2020).

Of course, the idea to audit software systems is not new. In fact, establishing procedures to ensure consistency with predefined requirements is a fundamental aspect of systems engineering (Leveson, 2011). Nevertheless, seminal papers by Sandvig et al. (2014) and Diakopoulos (2015) helped popularise the idea that automated decision-making systems (AMDS) should be audited with regards to not only their technical performance but also their alignment with ethical values. A rich and growing academic literature on EBA has since emerged, and a range of EBA procedures have been developed (see e.g., Cobbe et al., 2021; ForHumanity, 2021; Kazim et al., 2020b; Keyes et al., 2019; Zicari et al., 2021).[2]

EBA has also received much attention from policymakers and private companies alike. National regulators like the UK Information Commissioner's Office have provided guidance on how to audit ADMS (ICO, 2020), and professional services firms like Deloitte (2020), EY (2018), KPMG (2020), and PwC (2019) have all developed auditing (or 'assurance') procedures to help clients ensure that the ADMS they design and deploy are legal, ethical, and safe In short, a new industry focusing on EBA is already taking shape (Morley et al., 2021a).

Despite the surge in interest, important aspects of EBA – such as the feasibility and effectiveness of different auditing procedures – are yet to be substantiated by research. Raji and Buolamwini (2019) suggest that internal audits *can* help check that the engineering processes involved in designing ADMS meet specific standards. Similarly, Brundage et al. (2020) argue that external audits *can* help organisations verify claims about ADMS. These works have articulated important theoretical justifications for EBA. However, the affordances and constraints of EBA procedures can only be investigated and evaluated in applied contexts.

[1] Different researchers use different terms. LaBrie & Steinke (2019) call it 'ethical audits'. However, we prefer the term EBA to avoid any confusion: we do not refer to a kind of auditing done ethically, but to auditing procedures for which ethics principles constitute the baseline.
[2] For an overview of the literature on algorithmic audits, see Mökander et al. (2021b).

The literature on EBA contains few case studies: Buolamwini and Gebru (2018) assessed the efficacy of external audits to address biases in facial recognition systems; Mahajan et al. (2020) outlined a procedure to audit AI systems that replicate cognitive tasks in radiology workflows; and Kazim et al. (2021) applied a systematic audit to algorithmic recruitment systems. However, there is still little understanding of how organisations implement EBA and what challenges they face in the process. This article addresses that gap by providing insights from a longitudinal industry case study.

Over a period of 12 months, we observed and analysed the activities of AstraZeneca (a biopharmaceutical company) as it prepared for and underwent an ethics-based audit. This article describes and discusses the findings from that study to make two contributions to the existing literature. First, it provides a descriptive account of how a large, decentralised, and R&D-driven company like AstraZeneca implements AI governance in practice. Second, by outlining the challenges and tensions involved in conducting a real-world AI audit, it identifies transferable best practices for how to develop EBA procedures. Taken together, these contributions support the objective of outlining the conditions under which EBA is a feasible and effective mechanism for operationalising AI governance.

The findings from the case study suggest that the main difficulties organisations face when conducting EBA mirror corporate governance challenges. Organisations attempting to implement EBA must consider how to harmonise standards, demarcate the scope of the audit, define key performance indicators, and drive change management. These findings will not come as a surprise to management scholars. Yet efforts to operationalise AI governance are interdisciplinary in nature and the transfer of knowledge from different fields of study will be a key success factor to design and implement EBA procedures. This paper is thus aimed at computer scientists, ethicists, and auditors that develop EBA procedures as well as managers tasked with the implementation of corporate AI ethics principles.

The remainder of this article is structured as follows. Section3.2 draws on previous research to establish the need for corporate EBA. Section 3.3 introduces the case study by giving a descriptive account of AstraZeneca as an organisation as well as of the events leading up to the AI audit. Section 3.4 describes AstraZeneca's 2021 AI audit in greater detail, situating it relative to previous research on EBA. Section 3.5 describes the methodology used to conduct this study, which is based on participant observation and semi-structured interviews. Section 3.6 discusses the findings from the case study. Section 3.7 identifies limitations of the approach taken in this article. Finally, Section 3.8 highlights current best practices and directions for future research.

## 2. The need to operationalise AI governance

AI holds great promise to support human development and prosperity (Dignum, 2020). Enabled by advances in machine learning (ML), access to increased computing power, the growing availability of data, and the ubiquity of digital devices (Balas et al., 2020), AI systems can improve efficiency, reduce costs, and help solve complex problems (Taddeo et al., 2018).

The gains associated with AI technologies are not only economic but also social in nature. Take healthcare as an example. AI systems aid clinicians in medical diagnostics (Grote & Berens, 2020) and enable personalised treatments (Begoli et al., 2019). AI systems also drive healthcare service improvements through better forecasting (Kaushik et al., 2020). In the pharmaceutical industry the combination of pattern recognition for molecular structures and laboratory automation promises better and faster drug discovery and delivery processes (Schneider, 2019). In sum, using AI systems in the healthcare sector may allow humans to live more healthy lives while enabling societies to manage the rising costs associated with ageing populations (Jiang et al., 2017).

However, the use of AI systems in the healthcare sector is coupled with ethical challenges (Morley et al., 2020b; Topol, 2019). The use of AI systems may leave users vulnerable to discrimination and privacy violations (Leslie, 2019). It may also enable wrongdoing and erode human self-determination (Tsamados et al., 2020). Many of these risks apply to AI systems generally. But how AI systems process health data is particularly delicate (McLennan et al., 2022), since patients may be harmed by reputational damage and suboptimal care (Laurie et al., 2014). For example, recent studies have found racial biases in medical devices that provide pulse oximetry measurements (Sjoding et al., 2020).

While the adoption of AI systems has outpaced the development of governance mechanisms designed to address the associated ethical concerns (Taeihagh et al., 2021), abstaining from using AI systems in sensitive areas of application is not the way forward (Cookson, 2018). As far as the use of AI in medicine is concerned, a 'precautionary approach' would likely cause significant social opportunity costs due to constraints that undermine the development of promising technologies, drugs and treatments (Blasimme & Vayena, 2021). Moreover, AI systems inevitably are part of larger socio-technical systems that comprise other technical artefacts as well as human operators (Di Maio, 2014; van de Poel, 2020). No purely technical solution will thus be able to ensure that AI systems operate in ways that are ethically-sound (Lauer, 2020; Schneider et al., 2020).

It is essential that public and private actors seeking to benefit from AI systems understand and address the varied ethical challenges associated with their use. Responding to this need, numerous governments and NGOs have proposed ethical principles that provide normative guidance to organisations designing and deploying AI systems (Fjeld, 2020; Jobin et al., 2019).[3] These guidelines tend to converge on five principles: beneficence, non-maleficence, autonomy, justice, and explicability (Floridi & Cowls, 2019).[4] This is encouraging. Yet principles alone cannot guarantee that AI systems are designed and used in ethically-sound ways. The apparent consensus around normative principles hides deep political tensions around interpreting abstract concepts like fairness and justice (Mittelstadt, 2019). Moreover, translating principles into practice often requires trade-offs (Whittlestone et al., 2019). Most critically, the industry lacks useful tools to translate abstract principles into verifiable criteria (Vakkuri et al., 2019).

Due to these constraints, technology-oriented companies have struggled with operationalising AI ethics. Fortunately, companies need not start from scratch: numerous translational mechanisms for AI governance have been proposed and studied (Ayling & Chapman, 2021; Morley et al., 2021b). These include *impact assessments lists* (AI HLEG, 2020; Koshiyama, 2019; Reisman et al., 2018), *model cards* (Mitchell et al., 2019), *datasheets* (Gebru et al., 2018; Holland et al., 2018), as well as *human-in-the-loop* protocols (Jotterand & Bosco, 2020), *standards* and reporting guidelines for using AI systems (Cihon, 2019; Cruz Rivera et al., 2020; Liu et al., 2020), and the inclusion of broader *impact requirements* in software development processes (Prunkl et al., 2021).

All these efforts are complementary and serve the overarching purpose of enabling effective corporate AI governance. That is important because private companies significantly influence regulatory methods and technological developments (Cihon et al., 2021; Minkkinen et al., 2021). This dependency on private actors is a double-edged sword. On the one hand, competing interests can undermine even well-intentioned attempts to translate principles into practice (Floridi, 2019). On the other hand, private companies have strong incentives to implement effective AI governance to improve numerous business metrics like regulatory preparedness, data security, talent acquisition, reputational management, and process optimisation (EIU, 2020; Holweg et al., 2022).

[3] Recent and influential contributions include AI HLEG (2019), IEEE (2019) and OECD (2019).

[4] Healthcare practitioners will note the overlap here with the classical principles of bioethics (Dunn & Hope, 2018).

How AI systems are designed and used is a concern not only for individual organisations but also for society at large (Floridi, 2021). This insight has been reflected in recent regulatory developments. Both the EU Artificial Intelligence Act (AIA) (European Commission, 2021a) and the US Algorithmic Accountability Act of 2022 (Office of U.S. Senator Ron Wyden, 2022) constitute attempts to elaborate general legal frameworks for AI. Hard legislation can, if properly designed and enforced, address parts of the gap EBA procedures fill. For example, the AIA requires specific 'high-risk' AI systems to undergo so-called 'conformity assessments by the involvement of an independent third party' (Mökander et al., 2021a). But most AI systems are not classified as 'high-risk' and will thus not subject to the requirements stipulated in the AIA. Moreover, the use of AI systems may be problematic and deserving of scrutiny even when not illegal. In short, there will always be room for more and better, post-compliance, ethical behaviour (Floridi, 2018). The 'ethics-based' approach studied in this article is thus compatible with – and complementary to – hard legislation.

## 3. AstraZeneca and AI governance

AstraZeneca is a multinational biopharmaceutical company headquartered in Cambridge, UK. It has an annual turnover of $26bn and employs over 76,000 people (AstraZeneca, 2020a). As an R&D-driven organisation, AstraZeneca discovers and supplies innovative medicines worldwide. Its core business is using science and innovation to improve health outcomes through more effective treatment and prevention of complex diseases.[5] Recently, AstraZeneca has become a household name on account of the Oxford-AstraZeneca Covid-19 vaccine (Gilbert et al., 2021).

The biopharmaceutical industry has always been data-driven (Langkafel, 2015). To develop new treatments, researchers follow the scientific method by building and testing hypotheses about the safety and efficacy of various treatments.[6] For example, AstraZeneca relies heavily on statistical analysis to probe the efficacy of candidate drugs in the research pipeline. Hence, AstraZeneca has long-established processes for data, quality, and safety management. However, how data can be collected, analysed, and utilised keeps changing

[5] AstraZeneca is divided into three main therapy areas: Oncology; Cardiovascular, Renal and Metabolism; and Respiratory and Immunology diseases (AstraZeneca, 2021b).

[6] The process of discovering and developing new drugs is long and complex: only a small proportion of molecules that are identified as a candidate drug are approved (Ashenden, 2021).

(Ashenden et al., 2021). By harnessing the power of AI systems, researchers can find new correlations and draw useful inferences from the growing availability of data.

Examples of use cases of AI systems within AstraZeneca are abundant. For example, the company uses biological insight knowledge graphs (BIKG) to improve drug discovery and development processes (Crowe, 2020). Using BIKG helps synthesise, integrate, and leverage prior knowledge to gain new insights into disease characteristics and design smarter clinical trials (Vasetenkov, 2021). AI systems are also used for fast and accurate medical image analysis. Using AI systems based on ML and image recognition cuts analysis time by over 30% and improves accuracy (AstraZeneca, 2021a). Moreover, AI systems help automate various tasks. For example, AstraZeneca use natural language processing to prioritise adverse event reports (Lea et al., 2021; Rizk et al., 2021). Here, AI systems help classify events, separate outcomes by severity to enable appropriate action, thus leading to quicker response times and better patient experiences.

Despite excitement about these opportunities, AstraZeneca is conscious about risks associated with AI systems. As discussed in Section 3.2, these include concerns related to privacy, fairness, transparency and safety. In November 2020, AstraZeneca's board moved towards addressing these risks by publishing a set of Principles for Ethical Data and AI (henceforth, *ethics principles.* Se Table 3 below). These stipulate that the use of data and AI systems should be private and secure; explainable and transparent; fair; accountable; as well as human-centric and socially beneficial (AstraZeneca, 2020b).[7]

The primary aim of these *ethics principles* is to help employees and partners safely and effectively navigate the risks associated with AI systems.[8] However, for AstraZeneca, AI governance serves numerous additional purposes. To use AI systems in line with the overall company strategy helps realise synergies and maximise value creation. Moreover, the voluntary adoption of the *ethics principles* strengthens AstraZeneca's brand.[9] Finally, the same internal processes that allow AstraZeneca to demonstrate adherence to its *ethics principles* also help it manage legal risks by ensuring compliance with existing legislation and anticipating forthcoming legislation.

These advantages are potential and not guaranteed. Principles alone cannot ensure that AI systems are designed and used in ways that are ethical (Mittelstadt, 2019). Hence,

[7] The process of formulating AstraZeneca's *ethics principles* involved numerous internal workshops and consultations with external experts and stakeholders.
[8] The *ethics principles* are thus to be seen as an extension of AstraZeneca's overarching organisational values.
[9] As noted by Slee (2021), creating auditable algorithms and datasets is a promising avenue for organisations to bridge the presentation gab between brands and the AI systems they design and deploy.

AstraZeneca followed its commitment to its *ethics principles* by focusing on their implementation. However, doing so was not straightforward. AstraZeneca already had several related governance structures in place, e.g., with regard to quality and data management, corporate social responsibility (CSR), sustainability, and product safety. Furthermore, AstraZeneca is a decentralised organisation in which different business areas operate independently. This structure provides flexibility but complicates the agreement and enforcement of common standards and procedures. Taking those considerations into account, AstraZeneca allowed each business area to develop their own AI governance structures to reflect local variations in objectives, digital maturity, and ways of working – as long as these align with the externally published *ethics principles*.

To support local activities, however, four enterprise-wide initiatives were launched:

1) The creation of an overarching *compliance document*
2) The development of a *Responsible AI playbook*
3) The establishment of (i) an *AI resolution Board* and (ii) an internal *Responsible AI Consultancy Service*
4) The commissioning of an *AI audit* conducted by an independent party

First, a compliance document was created, breaking down each high-level principle into more tangible and actionable formulations. Table 2, below, illustrates how that document attempts to bridge the gap between principles and practice in AI ethics.

*Table 3. AstraZeneca's Principles for Ethical Data and AI usage*

| Principle | Operationalisation |
|---|---|
| Private and secure | We respect privacy and act in a manner compatible with intended data use |
| | We employ Data & AI Systems that are designed to be secure |
| Explainable and transparent | We are open about the use, strengths, and limitations of our AI systems |
| | We ensure assumptions are clear, algorithms are appropriately documented, decisions are explainable, and processes to manage unanticipated consequences |
| Fair | We endeavour to use robust, inclusive datasets in our Data & AI systems |
| | We treat people and communities fairly and equitably in the design, process, and outcome distribution of our AI systems |
| Accountable | We apply governance proportional to the impact and risk of AI systems |
| | We anticipate and mitigate the impact of potential unfavourable consequences of AI through testing, governance, and procedures |
| Human-centric and socially beneficial | Where Data & AI is involved, humans oversee the system and are accountable for driving clear, expected benefits to people and society |
| | We employ human-led governance over our AI systems. We respect human dignity and autonomy and strive to reflect this in our AI systems |

Second, a Responsible AI Playbook was developed to provide more detailed, end-to-end guidance on developing, testing, and deploying AI systems within AstraZeneca.[10] The Playbook is a continuously updated online repository directing AstraZeneca employees to relevant resources, guidelines and best practices. The Playbook also summarises the specific regulations applicable to different AI use cases.

Third, new organisational functions were established. Specifically, an AI resolution board was created to review 'high-risk' AI use cases and an internal Responsible AI Consultancy Service was launched to facilitate the sharing of best practices and to educate staff about the risks of using AI systems in different contexts. The Responsible AI Consultancy Service serves three objectives: providing ethical guidance; supporting the practical embedding of the *ethics principles*; and monitoring the governance of AI projects.

Fourth, and most relevant for our purposes, AstraZeneca underwent an AI audit. This audit constituted the research material for our case study, and framing it is the focus of the next section.

## 4. An 'ethics-based' AI audit

In Q4 2021, AstraZeneca underwent an AI audit. However, because the term 'AI audit' has been used in many different ways, some clarifications are needed to specify what we refer to in this case. The AI audit conducted within AstraZeneca was an ethics-based, process-oriented audit conducted in collaboration with an independent third party. The remainder of this section unpacks what this means in practice.

The audit was 'ethics-based' insofar as AstraZeneca's *ethics principles* constituted the baseline against which organisational practices were evaluated. In short, the audit concerned what ought to be done over and above existing legislation. Of course, AI audits can be employed by different stakeholders and for different purposes. For example, Brown et al.

[10] The Playbook was developed by AstraZeneca's R&D Data Office yet is accessible to everyone in the organisation.

(Brown et al., 2021) distinguish between AI audits used (i) by regulators to assess whether a specific system meets legal standards; (ii) by providers looking to mitigate risks; and (iii) by other stakeholders wishing to make informed decisions about how they engage with specific companies. The AI audit conducted within AstraZeneca corresponds to (ii) since it was directed towards demonstrating adherence to voluntary codes of conduct.[11]

Further, AstraZeneca's audit was 'external' because it involved the commissioning of an independent third-party auditor. Specifically, the audit was coordinated by AstraZeneca's internal audit function and conducted by an external service provider.[12] In the literature a distinction is often made between internal audits, based on self-assessment, and external audits conducted by expert organisations (Mantelero, 2018). The latter tend to be limited by reduced access to internal processes (Raji et al., 2020). However, involving external experts can address the confirmation bias that may prevent internal audits from recognising critical flaws (Bauer, 2016). By subjecting itself to external review, AstraZeneca thus got valuable feedback on how to improve its existing and emerging AI governance structures.[13]

A central idea underpinning EBA is that procedural regularity and transparency contribute to good governance (Floridi, 2017). Hence, one aim of EBA is to create traceable documentation.[14] However, transparency must always be understood in context, i.e., with regards to a specific audience and intended purpose (Larsson & Heintz, 2020). In AstraZeneca's case, the audit's audience was internal decision-makers, and its most obvious purpose was assessing the extent to which the *ethics principles* had been adopted.

Operationally, the audit conducted within AstraZeneca consisted of two types of activities: a high-level *governance audit* of organisational structures and processes and *in-depth audits* of specific projects that either develop or use AI systems. Here, it is worth mentioning that the subject matter of EBA can either be a process, an organisational unit, or a technical system.[15] These approaches are not mutually exclusive but rather crucially complementary (Mökander & Axente, 2021). AstraZeneca's AI audit focused on processes and people, i.e., on assessing (i) the soundness and completeness of organisational processes and

[11] Oversight is critical to operationalise AI governance. In practice, this implies establishing evidence of how the AI systems were created and how they are operating (Kroll, 2021).

[12] The company that conducted the AI audit is a leading professional services firm. In line with the non-disclosure agreement (NDA) for this research, its name is not disclosed. Instead, it is referred to as 'the external auditor'.

[13] Note that all the other three enterprise-wide activities conducted by AstraZeneca to operationalise AI governance (see Section 3.3) were internal in nature.

[14] As noted by Kroll (2021), public documentation serves its function when, and largely because, its creation forces organisations to consider how to develop systems that can be presented in the best possible light.

[15] A consequence of viewing AI systems as parts of larger sociotechnical systems is that AI governance concern not only technical artifacts but also the organisations that develop or operate these (Powers & Ganascia 2021).

(ii) the extent to which different organisational entities adhered to these processes. During technical audits, in contrast, AI systems' source codes can be reviewed (Mittelstadt, 2016) or, alternatively, the behaviour (i.e., outputs) of such systems can be tested for a wide range of different input values (Kroll et al., 2016). However, no technical audits of individual AI systems were conducted during AstraZeneca's AI audit.

In Section 3.6, we discuss lessons learned from this audit. However, qualitative findings are best interpreted in the context in which they emerged. Hence, before exploring the findings, the next section outlines how we collected and analysed the data.

## 5. Methodology: An industry case study

To investigate the feasibility and effectiveness of EBA, we adopted that 'pragmatic stance'[16] and conducted a *industry case study* (Bass et al., 2018; Yin, 1994). Specifically, we observed and analysed AstraZeneca's internal activities as it prepared for and underwent an AI audit. The case study was *longitudinal* (Thomson et al., 2003) insofar as it lasted 12 months.[17] Three research questions (RQs) guided our research:

- *RQ1: How do industry firms integrate EBA within existing governance structures?*
- *RQ2: What challenges do industry firms face when attempting to implement EBA in practice?*
- *RQ3: What are best practices for how to prepare for and conduct EBA?*

With respect to these RQs, AstraZeneca's AI audit constituted what Merton (Merton, 1987) calls 'strategic research material' for at least two reasons. First, as an organisation that regularly harnesses data and AI systems for process automation and R&D purposes, AstraZeneca had practical concerns that overlapped with the theoretical problems we sought to address. Second, the timing was advantageous, in that we could follow the entire journey, from the publication of AstraZeneca's *ethics principles* (in November 2020) to the evaluation of the AI audit during Q4 2021.

Methodologically, the present study leveraged two qualitative research methods: *participant observation* and *semi-structured interviews*. The former, in which research is carried out through the researcher's direct participation, has a long history in organisational research (Vinten, 1994). It is particularly well-suited to making sense of organisational

[16] According to the American pragmatists (notably C.S. Peirce, William James, and John Dewey) theories should be judged by their success when applied practically to real-world situations (Legg & Hookway, 2020).
[17] Longitudinal case studies have long been used to observe how different governance mechanisms impact organisational practices. See e.g., Jackall (2010)

practices (Woodside, 2016). In this study, participant observation meant embedding ourselves in the organisation to observe the activities associated with preparing for and conducting the AI audit. This involved joining weekly meetings, reviewing working documents, and taking notes – not only of the audit's eventual findings but also of the points raised and decisions made along the way.

Specifically, we observed two types of meetings: *internal meetings*, in which AstraZeneca employees prepared for (or evaluated the results from) the AI audit, and *audit meetings*, in which external auditors asked questions to, and reviewed documents provided by, AstraZeneca employees. Because AstraZeneca's employees are distributed internationally – and because of Covid-19-related travel restrictions – all meetings took place online.

In addition, we conducted semi-structured interviews (Edwards & Holland, 2013) with different stakeholders involved in the audit. This allowed us to follow up on themes emerging from regular audit meetings and explore different actors' motivations and perspectives.[18] In total, 18 people were interviewed – some on several occasions. Rather than interviewing a predefined list of people, a snowballing technique (Given, 2008) was used to recruit new interviewees. Nevertheless, we tried to interview representatives in different roles and strove for a balance between different genders, ethnicities, and educational backgrounds amongst the interviewees. Each interview lasted 1–2 hours. To make the participants feel comfortable and avoid disturbing the flow of meetings, we did not record interviews, taking notes instead.

NVivo was used to import, code,[19] and analyse meeting notes. A *parallel research design* (Creswell & Clark, 2011) was used, in which the participant observation and the semi-structured interviews were conducted simultaneously. This enabled an iterative process through which research findings could be triangulated (Frey, 2018), thereby minimising the risk of 'losing context' that is associated with qualitative coding (Bryman, 2016).

Ethical approval for this research was granted by the Oxford Internet Institute's departmental research ethics committee. Access to AstraZeneca's internal processes and stakeholders was obtained through an agreement that leveraged an existing institutional relationship: JM's doctoral research is funded through a studentship provided by AstraZeneca.[20] An NDA was signed with the external auditor, allowing us to join relevant meetings to study the process. In all meetings, participants were informed about our presence and the purpose of our research. No personal details were collected or stored.

[18] Another advantage of semi-structured interviews is that the conversation is directed to the problem under investigation rather than the researcher's preconceived interests (Wang & Yan 2012).
[19] Here, 'code' refers to a word that assigns an essence-capturing attribute for a phenomenon (Saldaña, 2009).
[20] The studentship is funded by AstraZeneca but administered and paid out by the University. There have been no direct financial transactions between AstraZeneca and myself. The research is academically independent, and all views expressed belong to the authors alone.

## 6. Lessons learned from AstraZeneca's 2021 AI audit

When analysing the data, we found that the answers to questions about how to best design and implement EBA procedures often hinge on decisions made earlier in the process of operationalising corporate AI governance. Hence, when presenting the findings, we start with high-level observations and proceed with increasing levels of specificity.

### *6.1 Balancing legitimate yet competing interests*

A fundamental tension exists between the need for risk management, on the one hand, and incentives for innovation on the other.[21] This tension is particularly acute for R&D-driven organisations like AstraZeneca – both from an ethical and a financial point of view. For example, when developing new treatments, it is essential to monitor patient responses from a safety perspective. Hence, AstraZeneca trains AI systems to detect treatment response patterns and associate biophysical reactions with the safety risks of specific drugs (Nadler et al., 2021). Excessive red tape could hamper the development and adoption of such, potentially lifesaving, procedures. This shows that it is often not possible to 'err on the safe side'. Both the pharmaceutical industry and society at large have an obligation to put patients' care and safety first – and this means using innovative technologies to develop new drugs as well as to diagnose and intervene as early as possible in the course of a disease.

Similarly, from a financial perspective, R&D-oriented activities always carry risks since they involve trying new ideas – which often fail to progress.[22] However, even 'failed' R&D projects inform pharmaceutical innovation (Chiou et al., 2012). Hence, risk per se is not undesirable from an organisational point of view. Rather, the priority is to define and control the risk appetite in different projects. From an auditing perspective, this dynamic has two direct implications. First, EBA procedures that duplicate existing governance structures, or are perceived as unnecessary, are unlikely to be feasible and effective. Second, post-hoc EBA procedures that only highlight the risks associated with specific AI use cases are less likely to be adopted than continuous EBA procedures that help technology providers define and regulate technology-related risks.

[21] Critically-oriented researchers often highlight AI systems' failures to stress the need for more regulation (Greene et al., 2019). In contrast, techno-optimists point towards the gains such systems bring and caution against red tape (Diamandis & Kotler, 2012).

[22] Pammolli et al. (2020) analysed R&D activities related to drug development and found that over 70% of projects initiated between 2000 and 2009 had been terminated within one year.

### *6.2 Demarcating the material scope for AI governance*

Another high-level observation concerns the difficulty to define the material scope of AI governance in general and EBA in particular. As is well-known, there is no universally accepted definition of AI (Wang, 2019).[23] Nevertheless, every policy needs to define its material scope (Schuett, 2019).[24] Consequently, when attempting to operationalise its *ethics principles*, AstraZeneca struggled to define the systems and processes to which they ought to apply. That is partly because both human decision-makers and AI systems have their own strengths and weaknesses (Baum, 2017) and partly because ethical tensions can sometimes be intrinsic to the decision-making tasks at hand (Danks & London, 2017).[25]

Within AstraZeneca, representatives from the internal audit function stressed that underinclusive definitions of AI may lead to potential risks going unnoticed and unmitigated. Other stakeholders, including some managers and statisticians from the IT and R&D departments, warned that overinclusive AI definitions risk adding unnecessary layers of governance to very well-established systems and processes. As one manager objected:

> *"We are not doing any AI projects. We are, of course, doing large scale analytics, but only using statistical techniques that have long been standard practice in the industry."* (P5)

To solve this tension, AstraZeneca did not try to define what AI *is*.[26] Instead, AstraZeneca's Responsible AI Playbook lists and exemplifies the functional capabilities of the systems to which their AI governance framework *applies*. For each functional capability (such as the ability to emulate cognitive tasks), the Playbook provides concrete examples. Amongst others, the Playbook states that statistical tests (e.g., a T-test) conducted during data analysis are outside AstraZeneca's AI governance framework's scope. In contrast, automated statistical tests informing decisions that impact humans (e.g., stratifying patients into different arms of a clinical trial) are within scope. A list of examples does not constitute a definition of AI, nor does it provide a sufficient basis on which to create an exhaustive inventory of an organisation's AI systems. Nevertheless, listing examples of use cases that are in (or out) of scope informs attempts to operationalise AI governance.

[23] Some researchers use the term AI to refer to a type of agents that display some levels of autonomy, adaptability, and problem-solving capacity (Legg & Hutter, 2007). Others take AI to demarcate the set of computational techniques designed to approximate cognitive tasks (US Defence Authorization Act, 2018). Yet others use the term to describe the science and engineering of making specific machines (McCarthy, 2007).
[24] See Mokander et al. (2022a) for a comparison of methods for classifying AI systems for governance purposes.
[25] As Bryson (2021) argues, problems associated with 'AI' have not so much been created as exposed by it.
[26] Within AstraZeneca a 'high-level' definition of AI exists. However, this definition is flexible enough to allow each business area to further refine the material scope of its AI governance activities.

Furthermore, AstraZeneca adopted a risk-based approach, whereby the level of governance required for a specific system is proportionate to its risk level.[27] This means that systems within scope are classified as either low-, medium- or high-risk, depending on (i) the types of risk the system poses to humans and the organisation and (ii) the extent to which it makes autonomous decisions without human judgement. The approach taken is pragmatic[28] since it enables managers and developers to determine whether the *ethics principles* apply to specific systems. At the same time, the approach makes it difficult to assemble an inventory of an organisation's various AI systems. Without such an inventory, AI auditors depended on the business to identify and select relevant projects and systems for the in-depth audits.

The main takeaway here is that designing and implementing EBA procedures is intrinsically linked to the question of material scope. Until the material scope of AI governance is accepted throughout the organisation, any EBA procedure would struggle to produce verifiable claims.

### *6.3 Harmonising standards across decentralised organisations*

A further challenge faced during the AI audit is rooted in the problem of ensuring harmonised standards across decentralised organisations. As mentioned, each business area within AstraZeneca operates independently. From an AI governance perspective, this implies that the business areas face different realities in terms of digital maturity, the type of AI systems employed, economic pressures, and employees' levels of training.

Consider the contrast between two functions within AstraZeneca: R&D and Commercial. First, there are operational differences. R&D routinely creates AI systems in-house to aid drug discovery and testing. An understanding of the statistical models underpinning different AI systems is therefore closely linked to R&D's core business. Within Commercial, sales representatives typically rely on data analytics software (like CSR systems or predictive modelling) as a means to an end. Second, there are structural differences. R&D relies on a centralised Data Office to manage and curate data. In contrast, analytics within Commercial is largely decentralised. Collaborating with external partners has many advantages, including the possibility to leverage local market knowledge and health data (see Section 6.5 below).

[27] A parallel can be made to the EU AIA, which also takes an explicitly risk-based approach to AI governance.

[28] Pragmatic problem-solving demands that things should be sorted so that their grouping will promote successful actions for some specific end (Dewey, 1957).

These operational and structural differences between business areas are reflected in their capacities to manage AI-related risks. Hence, different EBA procedures may be needed to assess each business area's governance structure.[29] For example, AstraZeneca's AI audit showed that business areas understood risk differently. Within R&D, many employees work directly with patients and patient data. Hence, they typically see patient-centric risks. As one of the interviewees stated:

> *"Some colleagues have been working with data protection for years. When they hear 'AI ethics', they immediately think of privacy breaches. I often have to remind them that AI ethics is more than just compliance with data protection laws."* (P2)

In contrast, employees working within the Commercial function typically understood risk in financial or contractual terms. Both perspectives are of course valid, and the only purpose of this example is to highlight the difficulty of harmonising a 'risk-based' approach across an organisation that encompasses different understandings of 'risk'.

However, this problem need not be insurmountable. A distinction is often made between *compliance assurance,* which aims at comparing a system to existing laws and regulations, and *risk assurance*, which corresponds to asking open-ended questions about how a system works (CDEI, 2021). Using this distinction, current best practice would demand harmonising EBA procedures that aim to provide compliance assurance across business areas. In contrast, EBA procedures that aim at risk assurance should be adapted locally to reflect how respective business areas understand risk.

### *6.4 Internal communication as a key to operationalising AI governance*

Our observations of AstraZeneca's AI audit suggest that internal communication and training efforts are central to operationalising corporate AI governance. These communication efforts were continuous and happened on several different levels in parallel. For example, the *ethics principles* were agreed upon through a bottom-up process that included extensive consultations with employees and external experts. Importantly, this process was not just about agreeing on a set of principles. It also aimed to anchor the proposed policy with key stakeholders internally. If, for example, managers and software developers do not understand or agree with a policy, they will not prioritise it. However, if they can see how it helps in their daily activities, they will likely adopt it even without top-down directives. As one AstraZeneca employee stated:

> *"Working with the AI Ethics and Governance team was beneficial as it pushed me to think about my project in different ways and gave me new points to consider when developing an AI solution."* (P17)

[29] Note that different EBA *procedures* does not imply different *objectives*. In AstraZeneca's case, the control objective of the audit was the same across all business areas whereas the method of verification varied due to the decentralised nature of the organisation.

Moreover, corporate AI governance is about change management. Having formulated the *ethics principles*, AstraZeneca proceeded to the implementation phase. To some extent, that required a time-consuming, top-down roll-out of value statements and compliance documents. This was not a straightforward task: employees have limited attention spans and are frequently bombarded with information about different governance initiatives. It took AstraZeneca over six months to formulate the principles and another year to embed them across the business. Even as the AI audit took place, pockets of the organisation remained unaware of the compliance document.

Previous academic literature has given much attention to (i) the principles that should guide the design and deployment of AI systems (Alshammari & Simpson, 2017; Floridi & Cowls, 2019) and (ii) the tools enabling managers and software developers to translate these principles into practice (Ayling & Chapman, 2021; Morley et al., 2020a). While these aspects remain important, our observations suggest that internal communication's role in corporate AI governance deserve more attention. After all, ensuring that AI systems are designed and used legally, ethically, and safely requires organisations to not only have the right values and tools in place but also to make their employees aware of them.

In terms of raising awareness, the findings suggest three best practices. First, communication concerning AI governance is most effective when supported by senior executives.[30] Second, communication efforts around specific EBA procedures work best when stressing how these are relevant to employees' daily tasks. Third, communication around EBA procedures should make explicit why these are needed, thus assuring staff that existing governance procedures are not being duplicated.

### *6.5 Upholding organisational values in procurement and external collaborations*

[30] This finding is supported by previous research. For example, Gasser and Schmitt (2021) have shown that the effectiveness of corporate governance mechanism depend on issues related to leadership, values and culture.

The full cycle of designing and deploying AI systems seldom takes place within one organisation. Typically, AI systems result from a complex and extended supply chain spanning a plurality of actors and different geographic regions (Crawford, 2021). For example, in 2019, AstraZeneca entered a strategic collaboration with the British start-up BenevolentAI to combine the former's scientific expertise and rich datasets with the latter's biomedical knowledge graph to better understand the mechanisms underlying chronic kidney disease and identify more efficacious treatments (BenevolentAI, 2019). Similarly, in 2021, AstraZeneca launched a collaboration with American healthcare company GRAIL to evaluate the effectiveness of early cancer detection technologies (GRAIL, 2021).

From a business perspective, external R&D collaborations offer numerous advantages.[31] However, such collaborations are coupled with several governance challenges. For example, AstraZeneca's compliance document stipulates that robust, inclusive datasets should be used to train AI systems. During the AI audit, the external auditors explored that by asking how datasets had been collected, cleaned, and processed. However, such EBA procedures are only effective in evaluating AI systems trained in-house. For systems procured from external vendors, neither AstraZeneca nor the independent auditors had full visibility of the internal processes of, or the data used by, suppliers and vendors when training these systems. When discussing the training data for a particular AI system, one participant in an audit meeting stated:

> *"I don't know to be honest. We don't have access to that data. I have tried to get access to the same data but without success. You will have to ask [the external partner]."* (P14)

This has several direct implications for EBA. First, to be effective, the same requirements must apply to all AI systems used by an organisation. Without harmonised requirements, there is a risk that potentially sensitive development projects will only be outsourced to external partners. Second, to be feasible, EBA procedures must encompass a review of corporate procurement processes. However, that may not necessarily require the creation of additional layers of governance. Rather, organisations should undertake a gap finding and filling exercise, adding ethics-based evaluation criteria to existing procurement processes.

### *6.6 Ethics-based auditing as a catalyst for internal change*

[31] External R&D collaborations benefit innovation by increasing efficiency, reducing costs, and granting access to valuable resources not available internally (Grimpe & Kaiser 2010).

There are many reasons why organisations subject themselves to EBA. For example, such audits can help to control technology-related risks and inform AI design choices. However, our observations suggest that there are also other motivations for conducting EBA. These include facilitating agenda setting, serving as a catalyst for internal change, and expanding organisational units' mandates.

First, AstraZeneca aims to leverage AI and other data-driven technologies to transform how research is conducted. Digitalisation has thus been put on top of the corporate agenda. However, as an organisation's technological resources evolve, old governance structures risk becoming ineffective. Hence, AstraZeneca has strong incentives to understand how its internal governance structures need to change to keep up with operational practices.

Second, while organisational change is often incremental, distinct events – such as an audit – can catalyse activities that increase the rate of change. Within AstraZeneca, the upcoming AI audit motivated managers to communicate with their teams about the *ethics principles* and incentivised business areas to develop appropriate governance mechanisms to demonstrate their adherence to those principles. Several interviewees even expressed concerns about how much focus was put on preparing for the audit as a discrete event:

> *"Whenever the upcoming audit took up too much of our internal focus, I felt the need to remind myself and the team that we are not trying to operationalise AI governance because of the audit but to do the right thing."* (P2)

Third, any governance initiative can expand the operational and budgetary mandates of specific organisational units. For example, depending on how AI governance initiatives are framed, they might extend the reach of central functions such as IT or increase the resources allocated to specific CSR initiatives. In AstraZeneca's case, the sustainability team drove the initial formulation of the *ethics principles*. Yet during roll-out, a more decentralised structure emerged, with each business area responsible for practically implementing the principles.

The point we seek to stress here is that identifying or mitigating harm resulting from AI failures is not the only reason to implement EBA. EBA procedures can – and often do – serve other important functions, e.g., catalysing organisational change.

### *6.7 Making verifiable claims on the basis of ethics-based audits*

The subject matter of AI audits can be a person, an organisation, a process, a system, or any combination thereof. The AI audit conducted within AstraZeneca took a process approach in which the assessment was based on management representation, e.g., through interviews with key decision-makers and a review of sample documentation. In line with this approach, no detailed reviews of source codes, data sets, or model outputs were performed. Some interviewees expressed surprise regarding this:

> *"We are only talking about basic assumptions and the completeness of our documentation. I don't see what this has to do with AI?"* (P14)

Despite some individuals' misgivings, the procedure followed during AstraZeneca's AI audit is well-supported by previous research. While AI systems may appear opaque, technologies can always be understood in terms of their designs and intended operational goals (Kroll, 2018). Similarly, third-party auditors can make verifiable claims about AI systems without accessing the underlying data and computational models by analysing publicly available information (Dash et al., 2019).

In fact, EBA procedures that focus on organisational processes have several advantages. They are less demanding than code audits in terms of access to proprietary data. Since proprietary protection is one of the main drivers of AI systems' opacity (Burrell, 2016; Pasquale, 2016), that facilitates the process of conducting AI audits. Moreover, EBA procedures focusing on organisational processes are explicitly forward-looking. Rather than conducting post-hoc evaluations, the auditor and the technology provider collaborate to assess and improve the processes that shape future AI systems' properties and safeguards. This helps distinguishing accountability from blame (Chopra & Singh, 2018).[32]

Nevertheless, it is important to remain realistic about what EBA procedures focusing on organisational processes can be expected to achieve. Such procedures can verify claims about technology providers' quality management systems but are fundamentally unable to produce verifiable claims about the impacts that autonomous, self-learning AI systems that co-evolve with complex environments may have over time.

### *6.8 Measuring progress and demonstrating success*

Social phenomena are increasingly measured, described, and influenced by numbers,[33] and the corporate governance field is no exception. Sine Taylor, management scholars have refined metrics to measure and control workers' productivity as well as the societal impact and environmental footprint of products and services (Cugueró-Escofet & Rosanas, 2017;

[32] According to Diakopoulos (2021), what is needed to operationalise AI governance in an organisation is a map that models the assignment of responsibility based on the ethical expectations of different actors.
[33] See Mau (2019)for an excellent account of the growing tendency to quantify the social world and how that process changes our assignment of worth.

Islam & Greenwood, 2021).[34] Such metrics are relevant for EBA for two reasons. First, organisations investing in corporate AI governance want to demonstrate success by pointing towards tangible improvements. Second, ethical decision-making requires a frame of reference, i.e., a baseline against which normative judgements can be made. EBA producers should, therefore, include metrics that quantify the behaviour of technology providers and the AI systems they design and deploy.

Recently, much literature has focused on measuring and assessing the performance of different AI systems along normative dimensions such as fairness, transparency, and accountability (Hoffmann et al., 2018). For example, Wachter et al. (Wachter et al., 2021) compiled a list containing over 20 different fairness metrics, accompanied by a guide for choosing the most appropriate one for different use cases. These metrics can, in turn, be leveraged by conceptual tools or software that measure, evaluate, or visualise one or more properties of AI systems during EBA (Bellamy et al., 2019; Cabrera et al., 2019).

However, the use of metrics during AI audits is not unproblematic. Goodheart's Law reminds us that when a measure becomes a target, it ceases to be a good metric (Greenfield, 2017). Moreover, as Lee et al. (Lee et al., 2021) argue, reductionist representations of normative values (like fairness) often bear little resemblance to how these notions are experienced in real-life. In practice, different principles often conflict and require trade-offs (Mittelstadt, 2019). Similarly, different definitions of fairness – like individual fairness and demographic parity – are mutually exclusive (Kusner et al., 2017; Verma & Rubin, 2018)

How suitable different metrics are for specific EBA procedures depends on the nature of the audit. For AstraZeneca's process audit, the metrics employed aimed at capturing the extent to which best practices within software development were followed and appropriate safeguards were in place. One way to do so would have been to record 'Yes'/'No' answers to simple checklists. Such an approach has some support; by formalising ad-hoc processes and empowering individual advocates, checklists help organisations identify risks and tensions (Madaio et al., 2020). Yet simply having a checklist is insufficient to ensure that AI systems are designed and used legally, ethically, and safely (McNamara et al., 2018) and previous research has found that checklists risk reducing auditing to a box-ticking exercise (Raji et al., 2020).

Rather than using binary checklists, the auditors in AstraZeneca's case made use of open-ended questions that allowed managers and developers to articulate how (and why) specific AI systems were built.[35] Indeed, the most fruitful moments happened when AstraZeneca's in-house experts and the external auditors jointly discussed the merits of different ways of measuring the properties of specific AI models – thereby challenging the assumptions that underpin concepts like fairness or transparency. For example, AstraZeneca staff were asked to consider questions like: do we have rules about when and how we use AI systems? and what evidence can we use to determine whether an AI systems we design is 'fair' or 'robust'? As one of the external auditors put it:

> *"The really rich information comes not from asking a pre-curated list of questions, but from listening to the answers and asking relevant follow-up questions."* (P9)

[34] Note that organisational performance metrics need not be based on financial measures alone. The perhaps most famous example of this is 'the balanced scorecard' (Kaplan & Norton, 1996).

Taken together, our observations before, during, and after AstraZeneca's AI audit suggest that the primary purpose of metrics in the process of operationalising AI governance is not to decide whether a specific system is 'ethical' or not, but rather to spark ethical deliberation, inform design choices, and help visualise the normative values embedded in that system. This observation is compatible with the claim that multi-dimensional Pareto frontiers can be used to strike publicly justifiable tradeoffs between competing criteria (Kearns & Roth, 2020). Thus, a fruitful avenue for future research would be to develop a guide on when and how to use different metrics in the software development lifecycle and as part of holistic EBA procedures.

### *6.9 The costs associated with ethics-based audits*

Efforts to operationalise AI governance inevitably incur both financial and administrative costs. In the case of EBA, that includes *initial costs* (e.g., time and resources invested in preparing for the audit as well as the procurement of audit services and test data) and *variable costs* (e.g., the costs of implementing and adhering to an audit's recommendations, such as additional steps in the development process or continuous human oversight).[36]

To start with, formulating organisational values bottom-up is a time-consuming activity. In AstraZeneca's case, the process of drafting and agreeing on the *ethics principles* included multiple consultations with executive leaders on strategy, with senior developers to understand AI-related risks, with heads of different business areas to compare the agreed-upon principles with existing codes of conduct, as well as with academic researchers and industry experts to receive external feedback. Subsequently, the *ethics principles* had to be communicated, anchored, and implemented across the organisation (another labour-intensive activity). Beyond the time invested by senior leaders and individual employees, approximately four full-time staff worked on driving and coordinating the implementation of AI governance within AstraZeneca during 2020 and 2021.

[35] Here, a parallel can be made to 'Ethical Foresight Analysis', a method based on Failure Modes and Effects Analysis (FMEA), which is standard practice in safety engineering (Floridi & Strait 2020).

[36] This is nothing new. Already in 1980, Weiss published an article titled *Auditability of Software: A Survey of Techniques and Costs*.

During the actual audit in Q4 2021, the demands on manual resources increased. A team of auditors were contracted to evaluate AstraZeneca's overarching AI governance structure and conduct in-depth reviews of selected AI development projects and use cases. The AI audit took 14 weeks to conduct. Throughout, AstraZeneca employees allocated time to provide the auditors with documents and answer detailed questions during interviews. Taken together, around 2,000 person-hours were invested in the audit, even though it was relatively light-touch and did not involve any technical tests of individual AI models.

These numbers only give a ballpark indication of the costs associated with EBA. Indeed, quantifying the costs associated with any governance mechanism is difficult. Take the ongoing debate concerning the costs of complying with the EU AIA as an example. According to the European Commission, obtaining certification for an AI system in line with the AIA will cost on average EUR 16,800–23,000, corresponding to approximately 10–14% of the development cost (Renda et al., 2021). While those numbers have been supported by independent researchers (Haataja & Bryson, 2021), the critics claim that the official estimates are too low and fail to incorporate the long-term effects of the legislation such as reduced investments in AI research (Mueller, 2021).

The discussion around the cost of complying with the EU AIA illustrates that a governance mechanism's financial viability does not hinge on its direct costs alone but also on long-term opportunity costs and transformative effects. After all, one of the main reasons why technology providers engage with auditors is that it is cheaper and easier to address system vulnerabilities early in the development process. For example, it can costs up to 15 times more to fix a software bug found during the testing phase than fixing the same bug found in the design phase (Dawson et al., 2010). This suggests that – despite the associated costs – businesses have clear incentives to design and implement effective EBA procedures.

## 7. Limitations

Conducting qualitative research is challenging and bound to result in methodological shortcomings (Miles & Huberman, 1994). Here, we discuss important limitations with regards to the validity, independence, and generalisability of the findings.

Consider validity first. Since this study relied on descriptive methods, it is most relevant to consider construct validity, i.e., the ability to link research observations to their intended theoretical constructs (Smith, 2014). For example, it is difficult to assess the ethical risks posed by specific AI systems. Therefore, we exclusively focused on *observing and describing* the challenges organisations face when implementing EBA procedures, rather than *identifying or measuring* the effects such procedures have on the behaviour of AI systems.

A further risk related to validity concerns the possibility of replicating findings from previous research due to confirmation bias (Pub et al., 2019). While difficult to eliminate, this risk was managed through an iterative process, with findings from the literature and the case study continuously informing each other. In fact, including longitudinal case studies helps strengthen the validity of nonexperimental research designs (Levendusky, 2013).

Another limitation concerns the independence of the research. As mentioned, JM's doctoral research is funded through an Oxford-AstraZeneca studentship. When such dependencies exist, researchers may feel pressured to produce 'positive' results, i.e., findings that the industry partner wants to hear (Maruyama & Ryan, 2014). To manage this risk, we communicated clear boundaries regarding our role as independent researchers. We also followed best practices in research ethics, e.g., informing all parties about the constructively critical nature of our work.

A final set of limitations concerns the generalisability of the case study's findings. Inevitably, the input provided by the industry partner can be biased or contextually limited (Morgan et al., 2016). Moreover, data controllers (like AstraZeneca) have an interest in not disclosing trade secrets (Flyvbjerg, 2001). We sought to reduce the risk that biased input distorts the analysis by triangulating the information provided by AstraZeneca employees with other sources. Still, the findings from the case study should not be treated as neutral, but rather as context-specific knowledge (Jackall, 2010).

These limitations do not mean that the findings cannot be generalised. Indeed, AstraZeneca's efforts to operationalise AI governance are highly representative of the many large firms that have recently adopted ethics principles for designing and deploying AI systems. Notable examples include BMW Group (2020), IBM (Cutler et al., 2018), Google (2018), and Microsoft (2019). In short, many large organisations will face similar challenges when developing and implementing EBA procedures. The findings presented in Section 3.6 will thus be relevant to other large corporations attempting to integrate EBA procedures within existing governance structures.

## 8. Conclusions

A new industry that focuses on auditing AI systems is emerging. The proposed European legislation on AI, which sketches the contours of a professionalised AI auditing ecosystem, is likely to accelerate this trend. In such a fast-moving and high-stakes environment, it is of increasing importance for both regulators and business executives to understand the conditions under which EBA is a feasible and effective mechanism for operationalising AI governance. The findings from the AstraZeneca case study helps further such an understanding.

Different EBA procedures serve different purposes. Process audits – such as that undertaken by AstraZeneca – are well suited to verifying claims about the quality management systems a particular technology provider has in place as well as to identifying, assessing, and mitigating risks throughout the AI life cycle. Compared to code audits, they are also less demanding in terms of access to proprietary information and sensitive data. However, it is important to remain realistic about process audits' capabilities and limitations. EBA procedures that do not include any technical elements are fundamentally unable to produce verifiable claims about the effects autonomous and self-learning AI systems may have over time.

In terms of implementation, our observations suggest that EBA procedures are most likely to be effective when integrated into existing governance structures. That is because EBA procedures that duplicate existing structures (or operate in silos) may be perceived as unnecessary by the managers and developers expected to implement them. Similarly, efforts to operationalise AI governance through EBA are most effective when internal communication is centred around how this would help employees with their daily tasks. In contrast, EBA procedures that are perceived as filling only abstract functions are easily reduced to box-ticking exercises – and thereby fail to positively influence the design and deployment of AI systems. Best practice thus demands that AI auditors – whether internal or external – collaborate with managers and software developers to counteract problems related to unethical uses of, and unforeseen risks posed by, AI systems.

Large multinational organisations attempting to operationalise AI governance through EBA will inevitably face at least three critical challenges. First, EBA's feasibility as a governance mechanism is undermined by the difficulty of harmonising standards across decentralised organisations. AI audits require a pre-defined baseline against which organisational units, processes, or systems can be evaluated. However, like AstraZeneca, large organisations often comprise distinct business areas operating independently. Mandating uniform AI governance structures top-down thus poses challenges to the entire way such organisations are structured and run.

Second, the lack of a well-defined material scope for AI governance constitutes an obstacle to EBA. In short, questions as to which systems and processes AI governance frameworks ought to apply to remain unanswered. AstraZeneca's difficulties when attempting to establish a material scope for their AI audit highlight the three-way trade-off between how precise a scope is, how easy it is to apply, and how generalisable it is. Nevertheless, pragmatic problem-solving demands that things should be sorted so that their grouping will promote successful actions for some specific end. As a result, it will remain difficult for any EBA procedure to produce verifiable claims until the material scope of AI governance is accepted throughout an organisation.

Third, unresolved tensions related to procurement and external R&D collaborations risk undermining AI audits' effectiveness. For example, to successfully operationalise AI governance, EBA procedures must treat AI systems developed in-house and those procured from third-party vendors equally. If not, new internal governance structures may cause unethical (or simply 'risky') development projects to be outsourced. This is akin to what Floridi (2019) has labelled 'ethics dumping', i.e., the malpractice of exporting unethical activities to countries (or organisations) where there are weaker legal and ethical frameworks and enforcement mechanisms. The solution here would be for organisations to include alignment with internal AI governance policies as a criterion in future procurement processes and contractual agreements with external R&D collaborators.

While the conclusions offered above may not be surprising, they nonetheless stand in contrast to what has hitherto been the focus of academic research in this field. Simplified, previous research on EBA fall into one of two categories. The first consists of works that draw on law and political or moral philosophy to justify why EBA is needed. The second consists of works that draw on computer science or systems engineering to specify how EBA ought to be conducted. However, both the best practices and the challenges highlighted in this article indicate that the main difficulties organisations face when conducting AI audits mirror classical governance challenges. This indicates that not only computer scientists, engineers, philosophers, and lawyers but also management scholars need to be involved in the research on how to design EBA procedures.[37]

[37] This conclusion reiterates findings from previous research. See e.g., Raisch & Krakowski (2021).

## APPENDIX 1

The following questions guided the semi-structured interviews we conducted with managers, software developers and internal auditors within AstraZeneca.

I did not plan to follow a manuscript but rather have open dialogues with the interviewees. Hence, the below questions only indicate the information we sought to extract from the interview module as a whole: the focus of individual interviews varied depending on the (professional) role and (personal) interests of the respective interview participant.

1. **Meta info**
   - 1.1. What is your job description/role within the organisation? (I.e., What are your primary responsibilities? What are your core tasks?)
   - 1.2. What are the major initiatives you are executing or planning to execute to achieve your goals and address your issues?
   - 1.3. How does your daily work relate to the design and deployment of AI systems?
   - 1.4. How (if at all) have you and your team been involved in drafting the AstraZeneca AI ethics principles and the internal AI governance framework?

2. **Context**
   - 2.1. How do (or would) you (and your team) define AI systems?
   - 2.2. Are you (or your team) developing or using AI systems to support your objectives?
   - 2.3. If yes, what are the primary benefits you hope to achieve through using or developing AI systems?
   - 2.4. What AI technologies are underpinning the application? I.e., predictive/diagnostic, symbolic/connectionist, fully automated/decision support etc.
   - 2.5. How probable, sensitive, and impactful are potential system failures for the application?
   - 2.6. What do you consider to be the most significant strengths of AI systems?
   - 2.7. What do you consider to be the biggest ethical risks posed by AI systems, both from an organisational and societal perspective?
   - 2.8. How do you see your department using AI in the next two years? What potential risks/governance factors do you think are relevant?

**3. AstraZeneca AI Ethics principles and governance**

3.1. How are you (and your team) currently managing ethical risks when using or developing AI systems? I.e., What are the existing AI and data governance processes, policies, methods, initiatives, and tools that can be used to ensure D&AI Ethics and AI governance? And how effective are they?
3.2. How do you think Ethics or D&AI Ethics applies to your usage of AI?
3.3. What would you consider as an important D&AI Ethics principle?
3.4. What is the value that such principles would bring to you (and to AstraZeneca)?
3.5. Who is accountable for decision-making regarding the design or usage of AI systems within your team? How do you ensure that the design and use of AI systems respect AstraZeneca's risk and compliance policies?
3.6. Is there a process for measuring data and algorithmic quality – biases, accuracy, balance, etc.? If yes, who is responsible for this?

**4. Ethics-based auditing**

4.1. Do you believe that Data & AI Ethics and AI Governance in general, and ethics-based auditing in particular, will benefit you and your team?
4.2. What is your role – and what are your responsibilities – within the emerging/recently implemented internal ethics-based auditing procedure?
4.3. What tools (e.g., software) and methods (e.g., assessment lists) are you using as part of the internal ethics-based auditing procedure?
4.4. Who/what is subject to the ethics-based audit; a person, an organisational unit, a software system or a technical component?
4.5. How does the process of ethics-based auditing of AI systems relate to existing structures of accountability and oversight?
4.6. What technical and practical constraints have you faced during the implementation of ethics-based auditing of AI systems?
4.7. How are you (and your team) managing these constraints associated with ethics-based auditing in practice?
4.8. What mechanisms incentivise the implementation of ethics-based auditing of AI systems for you and your team?
4.9. Is there a mechanism whereby the ethics-based auditing procedure is linked to personal accountability?

## 5. Suggestions for improvements

5.1. How stringent and enforceable should ethics-based auditing of AI systems be, in your opinion?

5.2. Which ethical key requirements do you think should be covered by ethics-based auditing procedures?

5.3. How do you think the AI systems you (and your team) are using should be evaluated and assessed? I.e., according to which methodology or metrics?

5.4. What is your recommendation to ensure ethics-based auditing is implemented successfully within AstraZeneca?